\documentclass[prb,aps,showpacs,twocolumn,floatfix,10pt]{revtex4-1}

\pdfoutput=1

\usepackage[english]{babel}

\usepackage[utf8]{inputenc}

\usepackage{amsmath,amssymb,amstext}

\usepackage{txfonts}

\usepackage{lmodern}

\usepackage{xcolor,soul}

\usepackage{bm} % for bold math symbols

\usepackage{listings} % for code

\usepackage{subfloat}

\usepackage{geometry} % to change the page dimensions
\usepackage{graphicx} % support the \includegraphics command and options

\usepackage{array} % for better arrays (eg matrices) in maths
\usepackage{paralist} % very flexible & customisable lists (eg. enumerate/itemize, etc.)
\usepackage{verbatim} % adds environment for commenting out blocks of text & for better verbatim
\usepackage{subfig} % make it possible to include more than one captioned figure/table in a single float
\usepackage[activate=normal]{pdfcprot} % optischer Randausgleich

\usepackage{fancyhdr} % This should be set AFTER setting up the page geometry
\usepackage[nottoc,notlof,notlot]{tocbibind} % Put the bibliography in the ToC
\usepackage[titles,subfigure]{tocloft} % Alter the style of the Table of Contents

\usepackage[font=small,labelfont=bf]{caption}
\usepackage{siunitx}
\usepackage{multirow}

\newcommand{\vect}[1]{\bm{#1}}
\newcommand*\diff{\mathop{}\!\mathrm{d}}

\newcommand*\pdiff{\mathop{}\!\mathrm{\partial}}

\newlength{\graphicswidth}
\newlength{\graphicswidthfull}
\definecolor{orange}{rgb}{1,0.5,0}
\definecolor{gray}{gray}{0.5}

\begin{document}

\title{Energy Conservation and Coupling Error Reduction\\in Non-Iterative Co-Simulations}
\author{Severin~Sadjina, Eilif~Pedersen%
%, Stian~Skjong
}
\affiliation{Department of Marine Technology, Norwegian University of Science and Technology, NO-7491 Trondheim, Norway}
%\author{Lars~Tandle~Kyllingstad}
%\affiliation{SINTEF Fisheries and Aquaculture, NO-7465 Trondheim, Norway}
%\date{} % Activate to display a given date or no date (if empty),
         % otherwise the current date is printed 

\begin{abstract}
When simulators are energetically coupled in a co-simulation, residual energies alter the total energy of the full coupled system.
This distorts the system dynamics, lowers the quality of the results, and can lead to instability.
By using power bonds to realize simulator coupling, the \emph{Energy-Conservation-based Co-Simulation} method (ECCO) [Sadjina \emph{et al}.\ 2016] exploits these concepts to define non-iterative global error estimation and adaptive step size control relying on coupling variable data alone.
Following similar argumentation, the \emph{Nearly Energy Preserving Coupling Element} (NEPCE) [Benedikt \emph{et al}.\ 2013] uses corrections to the simulator inputs to approximately ensure energy conservation.
Here, we discuss a modification to NEPCE for when direct feed-through is present in one of the coupled simulators.
We further demonstrate how accuracy and efficiency in non-iterative co-simulations are substantially enhanced when combining NEPCE with ECCO's adaptive step size controller.
A quarter car model with linear and nonlinear damping characteristics serves as a co-simulation benchmark, and we observe reductions of the coupling errors of up to \SI{98}{\percent} utilizing the concepts discussed here.
\end{abstract}

\maketitle

%------------------------------------------------------------ 
%------------------------------------------------------------ 
\section{Introduction}
\label{sec:introduction}

Co-simulation allows for the independent and parallel modeling and simulation of complex systems including multiple physical and engineering domains, the use of tailored software tools and expert knowledge, the efficient use of suited solvers, and the protection of intellectual property within models.
All these properties make this kind of simulator coupling an attractive choice, especially from an industrial perspective.
But the fact that coupled subsystems are solved independently of each other between discrete communication time points also emphasizes accuracy and stability issues.

The flow and the conservation of energy between simulators in a co-simulation can be conveniently studied when using \emph{power bonds} to realize the couplings.
A power bond is a direct energetic bond between subsystems defined by inputs and outputs whose product gives a physical power: force and velocity, electric current and voltage, pressure and flow rate, to name a few.
Because subsystems in a co-simulation advance in time independently of each other, energy transactions between them are inherently inaccurate.
\emph{Energy residuals} emerge as a consequence and directly affect the total energy of the overall coupled system.
Consequently, system dynamics are distorted and co-simulation accuracy and stability are challenged.

These concepts are exploited in the \emph{Energy-Conservation-based Co-Simulation} method\cite{Sadjina2016} (ECCO).
Because energy residuals are a direct expression of coupling errors, they are a versatile tool to assess the quality of co-simulations.
Based on such error estimators, ECCO defines an adaptive control of the co-simulation step size, and displays significant improvements in the accuracy and efficiency of non-iterative co-simulations.
Similar arguments are used in the \emph{Nearly Energy Preserving Coupling Element}\cite{Benedikt2013} (NEPCE) to introduce corrections to the flow of (generalized) power between simulators in order to minimize coupling errors.

Here, we have a closer look at NEPCE and its energy-conserving properties.
We further propose a modification to include the presence of direct feed-through, enhancing its performance.
NEPCE's efficiency is based on the assumption that the coupling variables are slowly varying functions of time.
This assumption is challenged, however, by finding a suitable choice of the co-simulation (macro) time step.
We demonstrate how this issue is efficiently handled by ECCO's energy-conservation-based adaptive step size control in order to substantially improve accuracy and efficiency.
Because the resulting framework is non-iterative, it is computationally inexpensive and well suited for industrial applications.

This paper is organized as follows:
In Section~\ref{sec:energy_conservation_in_co-simulations}, we start with a brief recapitulation of the flow and conservation of energy in co-simulations using power bonds.
Next, we study NEPCE's non-iterative corrections to the simulator inputs in Section~\ref{sec:energy_conserving_corrections} and show how they should be modified in the presence of direct feed-through.
Section~\ref{sec:adaptive_step_size} discusses how these corrections can be combined with ECCO's adaptive step size control, and a quarter car model is then used in Section~\ref{sec:co-simulation_benchmark_tests} to demonstrate the performance of the proposed method and its influence on co-simulation accuracy and efficiency.
Finally, we give a conclusion in Section~\ref{sec:conclusion}.

%------------------------------------------------------------ 
%------------------------------------------------------------ 
\section{Energy Conservation in Co-Simulations}
\label{sec:energy_conservation_in_co-simulations}

Most commonly, co-simulations are realized by letting the simulators advance in time in parallel and independently of each other, and then synchronizing coupling data at discrete communication time points.
This weak coupling approach is easily implemented and relatively efficient on paper:
It is universally applicable for industrial applications (which usually prohibit iterative schemes) and the parallelization potential holds the promise of substantial simulation speed-ups.
Its major weaknesses, however, are accuracy and stability.
Input quantities are generally unknown during the time integrations inside the simulators.
They must therefore be approximated, and are often simply held constant.
A sufficiently small macro time step has to be chosen in order to keep the coupling errors which result from this scheme contained.

\begin{figure}[h!tb]
	\centering
	\def\svgwidth{\graphicswidth}
	\subfloat[Inputs are set at $t = t_i$]{
		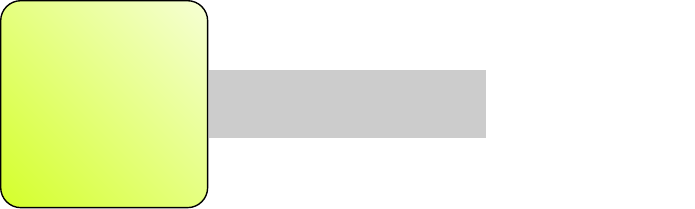
		\label{fig:power_bond_cosim_1}
	}\\
	\def\svgwidth{\graphicswidth}
	\subfloat[Outputs are retrieved at $t = t_{i+1}$]{
		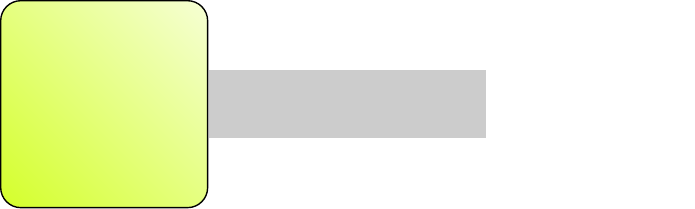
		\label{fig:power_bond_cosim_2}
	}
	\caption{%
		Two coupled simulators exchange energy through a power bond in a co-simulation
	}
	\label{fig:power_bond_cosim}
\end{figure}

%------------------------------------------------------------ 
\subsection{Power and Energy Residuals}
\label{subsec:residuals}

\begin{figure}[h!tb]
	\centering
	\def\svgwidth{\graphicswidth}
	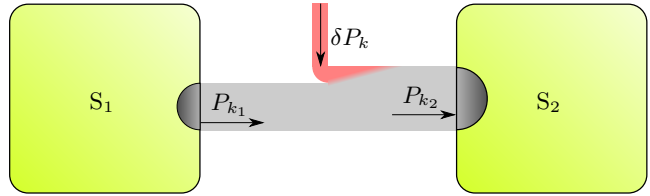
	\caption{%
		Total system dynamics are distorted by a residual power $\delta P_k$ between two energetically coupled simulators due to the independent time integrations
	}
	\label{fig:power_bond_residual}
\end{figure}

The use of power bonds from bond graph theory\cite{Paynter1961,Breedveld1984} allows to reframe these issues in terms of energy conservation considerations.\cite{Sadjina2016}
A power bond $k$ is defined by a pair of power variables---a flow $f_k$ and an effort $e_k$---whose product $P_k = e_k f_k$ gives a physical power.
Powers and energies, the universal currencies of physical systems, are directly accessible in co-simulations when using power bonds.
As an example, consider the flow of energy between two simulators S$_1$ and S$_2$ that are coupled via a power bond $k$, see Fig.~\ref{fig:power_bond_cosim}.
From the point of view of S$_1$, energy is transferred to S$_2$ at a rate
\begin{subequations}
\label{equ:energyflow}
\begin{equation}
\label{equ:energyflow_S1}
	P_{k_1}(t)
	=
	\mbox{$\tilde{u}$}_{k_1}(t)
	y_{k_1}(t)
	,
\end{equation}
where $y_{k_1}(t)$ is the output and $\mbox{$\tilde{u}$}_{k_1}(t) \approx u_{k_1}(t)$ is an approximation of the generally unknown value $u_{k_1}(t)$.
If, instead, we consider the energy transfer from the other simulator's perspective, we conclude that
\begin{equation}
\label{equ:energyflow_S2}
	P_{k_2}(t)
	=
	\mbox{$\tilde{u}$}_{k_2}(t)
	y_{k_2}(t)
	.
\end{equation}
\end{subequations}
This is problematic because it fundamentally violates the conservation of energy,
\begin{equation}
\label{equ:energyflow_dP}
	-
	(
	P_{k_1}
	+
	P_{k_2}
	)
	\neq
	0
	,
\end{equation}
because, generally, $\mbox{$\tilde{u}$}_{k_1}(t) \neq u_{k_1}(t)$ and $\mbox{$\tilde{u}$}_{k_2}(t) \neq u_{k_2}(t)$.
Hence, a \emph{residual energy} is incorrectly created due to the independent time integrations of the simulators during the macro time step $t_i \rightarrow t_{i+1} = t_i + \Delta t_i$,\cite{Sadjina2016}
\begin{subequations}
\label{equ:residuals}
\begin{align}
\label{equ:residuals_energy}
	\delta E_k(t_{i+1})
	&\equiv
	\int_{t_i}^{t_{i+1}}
	\delta P_k(t)
	\diff t
	,
\intertext{where}
\label{equ:residuals_power}
	\delta P_k
	&\equiv
	-
	(
	P_{k_1}
	+
	P_{k_2}
	)
\end{align}
\end{subequations}
is the \emph{residual power} for the power bond $k$, see Fig.~\ref{fig:power_bond_residual} for an illustration.
At each macro time step, the residual energy $\delta E_k$ is directly added to the total energy of the overall coupled system.\cite{Sadjina2016}
As a consequence, system dynamics are distorted and the quality of the co-simulation reduced.
Note that the power transmitted from S$_1$ to S$_2$ can be obtained from the simulator outputs as
\begin{equation}
\label{equ:energytransfer}
	P_{k_{12}}(t)
	=
	\sigma_{k_{12}}
	\big(
		y_{k_1}(t)
		y_{k_2}(t)
	\big)
	,
\end{equation}
where the sign $\sigma_{k_{12}} \equiv ({\vect{L}_k}_{12}-{\vect{L}_k}_{21})/2$ is determined by the corresponding elements of the connection graph matrix $\vect{L}$.

Luckily, inaccurate energy transactions provide us with a versatile error estimator because the corresponding residual energies are a direct expression of the co-simulation coupling errors and the violation of energy conservation.\cite{Sadjina2016}
This is exploited by the ECCO algorithm to define an adaptive macro step size controller:
For input extrapolation of order $m$, the residual energy scales quadratic with the step size\cite{Sadjina2016}, $\delta {E_k} = \mathcal{O}({\Delta t}^{m+2})$.
Consequently, the conservation of energy can be approximately satisfied by controlling the macro step size, optimizing the quality and efficiency of co-simulations.

%------------------------------------------------------------ 
\subsection{Local Errors in the Coupling Variables}
\label{subsec:errors}

Considering the time evolution of the internal states $\vect{x} = \{ x_1, x_2\}$ of the coupled simulators between the discrete communication time points $t_i$ and $t_{i+1}$,
\begin{subequations}
\label{equ:cosimulation}
\begin{equation}
\label{equ:cosimulation_evolution}
	\dot{\vect{x}}(t)
	=
	\vect{f}
	\big(
		\vect{x}(t)
		,
		\tilde{\vect{u}}(t)
	\big)
	,
	\quad
	t \in (t_i,t_{i+1}]
	,
\end{equation}
simulator coupling can be expressed as
\begin{align}
\label{equ:cosimulation_outputs}
	\vect{y}(t_{i+1})
	&=
	\vect{g}
	\big(
		\vect{x}(t_{i+1})
		,
		\tilde{\vect{u}}(t_{i+1})
	\big)
	,
	\\
\label{equ:cosimulation_connections}
	\vect{u}(t_{i+1})
	&=
	\vect{L}
	\vect{y}(t_{i+1})
	,
\end{align}
\end{subequations}
where $\vect{L}$ is a connection graph matrix that relates outputs $\vect{y}$ and inputs $\vect{u}$ at communication time points.
In the non-iterative co-simulation, the inputs are generally unknown and have to be approximated during the time integrations inside the simulators, $\tilde{\vect{u}}(t) \approx \vect{u}(t)$.
Most commonly, they are simply held constant such that $\tilde{\vect{u}}(t) = \vect{u}(t_i)$ for $t \in (t_i,t_{i+1}]$.

Let us in the following have a closer look at the local coupling errors which stem from these approximations and the independent time integrations in the subsystems between communication time points.
For the case of coupling via power bonds, these errors are conveniently represented as power and energy errors and directly related to the conservation of energy throughout the entire coupled system.
Using energies and powers as error metrics instead of non-energetic quantities has two major advantages:
\romannumeral 1.) They offer a more holistic and intuitive approach by considering the flow of energy between subsystems directly 
\romannumeral 2.) They avoid that some simulator's contributions to the global error are given too much weight.
If, for example, one simulator outputs a force and another a position, the numerical values of the force output will typically be much larger than those of the position output.
The same will then generally be true for the numerical values of the corresponding errors, skewing the actual simulators' contributions to the global co-simulation error.
The use of energy and power errors solves this issue in an elegant fashion.

In the next section, we will discuss how we can minimize local coupling errors.
The subsystem states are inaccessible in a typical co-simulation setting and can not be directly altered.
Instead, corrections to the inputs can be derived such that the residual energies between simulators are minimized and energy conservation is approximately satisfied.
These corrections ideally cancel the local errors in the inputs which are given by
\begin{subequations}
\label{equ:errors}
\begin{equation}
\label{equ:errors_input_general}
\begin{split}
	\Delta \vect{u}(t)
	&=
	\tilde{\vect{u}}(t)
	-
	\vect{u}_0(t)
	\\
	&=
	\tilde{\vect{u}}(t)
	-
	\vect{L}
	\vect{y}_0(t)
	\\
	&=
	\tilde{\vect{u}}(t)
	-
	\vect{L}
	\big(
		\vect{y}(t)
		-
		\Delta \vect{y}(t)
	\big)
	,
\end{split}
\end{equation}
where $\vect{u}_0(t)$ is the exact solution and we used that $\vect{u}_0(t)=\vect{L}\vect{y}_0(t)$ for any time $t$.
The errors in the outputs evaluate to
\begin{equation}
\label{equ:errors_output}
\begin{split}
	\Delta \vect{y}(t)
	&=
	\vect{y}(t)
	-
	\vect{y}_0(t)
	\\
	&=
	\vect{g}
	\big(
		\vect{x}(t),
		\tilde{\vect{u}}(t)
	\big)
	-
	\vect{g}
	\big(
		\vect{x}_0(t),
		\vect{u}_0(t)
	\big)
	\\
	&=
	\vect{J}_{\vect{g}}(\vect{u})
	\Delta \vect{u}(t)
	+
	\vect{J}_{\vect{g}}(\vect{x})
	\Delta \vect{x}(t)
	\\
	&+
	\mathcal{O}({\Delta t}^{m+2})
	,
\end{split}
\end{equation}
where ${J_{\vect{g}}}_{i j}(\vect{u}) = \pdiff g_i /\pdiff u_j$ is the interface Jacobian and ${J_{\vect{g}}}_{i j}(\vect{x}) = \pdiff g_i /\pdiff x_j$.
While the error contributions from the state vector are
\begin{equation}
\label{equ:errors_state}
	\Delta \vect{x}(t)
	=
	\vect{x}(t)
	-
	\vect{x}_0(t)
	=
	\mathcal{O}({\Delta t}^{m+2})
	,
\end{equation}
\end{subequations}
the input errors appear to order $\Delta \vect{u} = \mathcal{O}({\Delta t}^{m+1})$.
Consequently, if one of the simulators S$_k$ has direct feed-through, the output errors are also of order $\mathcal{O}({\Delta t}^{m+1})$ because then ${J_{\vect{g}}}_{k k}(\vect{u}) \neq 0$.
Using Eq.~\eqref{equ:errors_output} in Eq.~\eqref{equ:errors_input_general} and rewriting thus gives
\begin{equation}
\label{equ:errors_input}
\begin{split}
	\Delta \vect{u}(t)
	&=
	\big(
		1
		-
		\vect{L}
		\vect{J}
	\big)^{-1}
	\big(
		\tilde{\vect{u}}(t)
		-
		\vect{L}
		\vect{y}(t)
	\big)
	\\
	&+
	\mathcal{O}({\Delta t}^{m+2})
	,
\end{split}
\end{equation}
where we set $\vect{J} \equiv \vect{J}_{\vect{g}}(\vect{u})$ for brevity.
%
%Let us now have a look at the energy errors.
%The local error in the power transmitted through the power port $k_1$ of simulator S$_1$ is given by
%\begin{equation}
%\label{equ:errors_power_general}
%\begin{split}
%	\Delta P_1(t)
%	&=
%	P_1(t)
%	-
%	P^0_1(t)
%	\\
%	&=
%	\mbox{$\tilde{u}$}_1(t)
%	y_1(t)
%	-
%	u_1^0(t)
%	y_1^0(t)
%	\\
%	&=
%	\mbox{$\tilde{u}$}_1(t)
%	y_1(t)
%	-
%	L_{1 2}
%	y_1^0(t)
%	y_2^0(t)
%	\\
%	&=
%	\mbox{$\tilde{u}$}_1(t)
%	y_1(t)
%	-
%	L_{1 2}
%	y_1(t)
%	y_2(t)
%	\\
%	&+
%	L_{1 2}
%	\big(
%		y_1(t)
%		\Delta y_2(t)
%		+
%		y_2(t)
%		\Delta y_1(t)
%	\big)
%	\\
%	&+
%	\mathcal{O}({\Delta t}^{m+2})
%	.
%\end{split}
%\end{equation}
%Consequently, in the absence of direct feed-through
%\begin{subequations}
%\label{equ:errors_power}
%\begin{equation}
%\label{equ:errors_power_P1}
%\begin{split}
%	\delta P_1(t)
%	&\equiv
%	-\Delta P_1(t)
%	\\
%	&=
%	L_{1 2}
%	y_1(t)
%	y_2(t)
%	-
%	\mbox{$\tilde{u}$}_1(t)
%	y_1(t)
%	\\
%	&+
%	\mathcal{O}({\Delta t}^{m+2})
%\end{split}
%\end{equation}
%and, analogously,
%\begin{equation}
%\label{equ:errors_power_P2}
%\begin{split}
%	\delta P_2(t)
%	&\equiv
%	-\Delta P_2(t)
%	\\
%	&=
%	L_{2 1}
%	y_1(t)
%	y_2(t)
%	-
%	\mbox{$\tilde{u}$}_2(t)
%	y_2(t)
%	\\
%	&+
%	\mathcal{O}({\Delta t}^{m+2})
%	.
%\end{split}
%\end{equation}
%\end{subequations}

%------------------------------------------------------------ 
%------------------------------------------------------------ 
\section{Non-Iterative Energy-Conservation-Based Corrections}
\label{sec:energy_conserving_corrections}

Let us now take the idea of energy conservation in co-simulations a step further by directly modifying the coupling variables such that energy transactions between simulators are described more accurately.
In this section, we will explore this concept which is used by NEPCE~\cite{Benedikt2013} and generalize it to include the presence of direct feed-through.
In Section~\ref{sec:adaptive_step_size} we then discuss how the energy-conservation-based corrections studied here can be combined with ECCO's non-iterative adaptive step size controller, and Section~\ref{sec:co-simulation_benchmark_tests} demonstrates the substantial improvements in accuracy and efficiency thus obtained using a quarter car co-simulation benchmark model.

As can be seen from Eqs.~\eqref{equ:energyflow} and~\eqref{equ:residuals}, a residual energy
\begin{equation}
\label{equ:residual_energy_general}
	\delta E_k(t_{i+1})
	=
	-
	\int_{t_i}^{t_{i+1}}
	\tilde{\vect{u}}_k(t)
	\cdot
	\vect{y}_k(t)
	\diff t
\end{equation}
is accumulated during the time step $t_i\rightarrow t_{i+1}$ for a power bond $k$ connecting the inputs $\tilde{\vect{u}}_k = \{\mbox{$\tilde{u}$}_{k_1},\mbox{$\tilde{u}$}_{k_2}\}$ and outputs $\vect{y}_k = \{y_{k_1},y_{k_2}\}$.
The concept behind NEPCE is to find corrections $\delta \vect{u}_k = \{\delta u_{k_1}, \delta u_{k_2}\}$ to the inputs at communication time instant $t=t_i$ with the aim of reducing the residual energy by a factor of $(1-\alpha)$, such that
\begin{equation}
\label{equ:residual_energy_corrections}
	(\alpha-1)
	\delta E_k(t_{i+1})
	=
	\int_{t_i}^{t_{i+1}}
	\big(
		\tilde{\vect{u}}_k(t)
		+
		\delta \vect{u}_k(t)
	\big)
	\cdot
	\vect{y}_k(t)
	\diff t
\end{equation}
with the tuning factor $\alpha \in [0,1]$.
Ideally, $\alpha = 1$ if the corrections accurately track the errors in the inputs, $\delta \vect{u}_k(t) = - \Delta \vect{u}_k(t)$.
While this can not be realized in general for non-iterative co-simulations, however, corrections should be of the same order as the errors in the input~\eqref{equ:errors_input_general}, $\delta \vect{u}_k = \mathcal{O}({\Delta t}^{m+1})$, to mitigate their effects.
Moreover, a correction to the input will generally elicit a modification of the output $\delta \vect{y}_k$, such that we generally need to consider
\begin{equation}
\label{equ:residual_energy_corrections_feedthrough}
\begin{split}
	&(\alpha-1)
	\delta E_k(t_{i+1})
	\\
	=
	&\int_{t_i}^{t_{i+1}}
	\big(
		\tilde{\vect{u}}_k(t)
		+
		\delta \vect{u}_k(t)
	\big)
	\cdot
	\big(
		\vect{y}_k(t)
		+
		\delta \vect{y}_k(t)
	\big)		
	\diff t
	.
\end{split}
\end{equation}
If direct feed-through is present, this modification to the output is of the same order as the input corrections, $\delta \vect{y}_k = \mathcal{O}({\Delta t}^{m+1})$, and should be included.

%------------------------------------------------------------ 
\subsection{NEPCE}
\label{subsec:corrections_NEPCE}

But first, let us discuss the case where none of the simulators have direct feed-through.
Then, the errors in the inputs~\eqref{equ:errors_input} are simply 
\begin{equation}
	\Delta \vect{u}(t)
	=
	\tilde{\vect{u}}(t)
	-
	\vect{L}
	\vect{y}(t)
	+
	\mathcal{O}({\Delta t}^{m+2})
	,
\end{equation}
and Eq.~\eqref{equ:residual_energy_corrections} suffices.
Choosing
\begin{equation*}
	\delta \vect{u}(t) = -\Delta \vect{u}(t) \approx \vect{L} \vect{y}(t) - \tilde{\vect{u}}(t)
\end{equation*}
would make the residual energy vanish and the coupling quantities exact to order $\mathcal{O}({\Delta t}^{m+1})$.
As already mentioned, this is not possible for non-iterative co-simulations because $\vect{y}(t)$ is unknown a priori for $t = (t_i, t_{i+1}]$.
Instead, we realize the correction in terms of previous coupling data,\cite{Benedikt2013}
\begin{equation}
\label{equ:corrections_NEPCE}
	\delta \vect{u}(t)
	\approx
	\frac{\alpha}{\Delta t_i}
	\int_{t_{i-1}}^{t_i}
	\big(
		\vect{L}
		\vect{y}(\tau)
		-
		\tilde{\vect{u}}(\tau)
	\big)		
	\diff \tau
	,
\end{equation}
for $t \in (t_i,t_{i+1}]$, assuming that the coupling variables and the errors are slowly varying on the scale of the time step $\Delta t$.

Note that this is a reasonable assumption in theory:
In a co-simulation the macro time step should be chosen such that the dynamics of the system can be sufficiently well resolved in time.
A violation of this assumption is equivalent to the macro time step simply being too large for the problem at hand.
In section~\ref{sec:adaptive_step_size}, we will take a big step towards ensuring that this crucial assumption holds by combining the energy-conservation-based input corrections discussed in the present section with the energy-conservation-based adaptive step size controller ECCO.

%------------------------------------------------------------ 
\subsection{Corrections with Direct Feed-Through}
\label{subsec:corrections_feedthrough}

As discussed previously, corrections to the inputs cause modifications to the outputs which are of the same order $\mathcal{O}({\Delta t}^{m+1})$ in the presence of direct feed-through.
The errors in the inputs are then given by Eq.~\eqref{equ:errors_input}, and Eq.~\eqref{equ:corrections_NEPCE} should be modified to
\begin{equation}
\label{equ:corrections_feedthrough}
	\delta \vect{u}(t)
	\approx
	\frac{\alpha}{\Delta t_i}
	\big(
		1
		-
		\vect{L}
		\vect{J}
	\big)^{-1}
	\int_{t_{i-1}}^{t_i}
	\big(
		\vect{L}
		\vect{y}(\tau)
		-
		\tilde{\vect{u}}(\tau)
	\big)		
	\diff \tau
\end{equation}
to include all coupling errors of order $\mathcal{O}({\Delta t}^{m+1})$.
It is important to point out that Eq.~\eqref{equ:corrections_feedthrough} requires the knowledge of the generally time-dependent interface Jacobian $J_{i j} = \pdiff g_i /\pdiff u_j$.
In practical applications, it will likely not be available and the unmodified NEPCE form~\eqref{equ:corrections_NEPCE} should be chosen.
While disregarding the output error contribution in Eqs.~\eqref{equ:errors}, it is still an improvement over the uncorrected co-simulation in the presence of direct feed-through.

Finally, note that we can safely disregard the case where both simulators have direct feed-through, because it amounts to an algebraic loop which indicates that the particular system reticulation is not suitable for non-iterative co-simulation and ill-chosen.

%------------------------------------------------------------ 
%------------------------------------------------------------ 
\section{Energy-Conserving Adaptive Step Size Control}
\label{sec:adaptive_step_size}

The previous section discussed NEPCE and how it should be modified in the presence of direct feed-through.
The approach to energy-conservation-based corrections to the inputs in non-iterative co-simulations relies on the assumption that the coupling variables are slowly varying functions of time on the scale of the macro time step.
When this assumption does not hold the corrections become increasingly ineffective and can even lead to instability by exciting relatively fast dynamics in the subsystems\cite{Benedikt2013}.
In other words, the smaller the chosen macro time step the more efficient and beneficial the input corrections become.

The \emph{Energy-Conservation-based Co-Simulation} method (ECCO) provides a framework that allows us to adaptively choose a macro step size which (given some tolerances) approximately ensures energy conservation in non-iterative co-simulations.
This concept and its performance have recently been studied\cite{Sadjina2016}, and we shall in the following combine it with the energy-conservation-based input corrections from the previous section to define a non-iterative co-simulation framework yielding high accuracy and efficiency without the use of any simulator-internal data.

An I-controller is used to determine a new optimal step size
\begin{equation}
\label{equ:step_control}
	{\Delta t}_{i+1}
	=
	\alpha_\text{s}
	\epsilon(t_i)^{-k_\text{I}}
	\Delta t_i
\end{equation}
as a function of an error indicator $\epsilon$.
Here, $k_\text{I} = 0.3/(m+2)$ is the integral gain\footnote{%
	The denominator $m+2$ represents the order of the error, here $\delta E_k = \mathcal{O}({\Delta t}^{m+2})$.
	The corrections discussed in Sec.~\ref{sec:energy_conserving_corrections} have the aim of canceling the energy error $\delta E_k$ to leading order in $\Delta t$, and thus one should choose $k_\text{I} = 0.3/(m+3)$.
	Here, we decide against this alteration for two reasons:
	\romannumeral 1.) As mentioned previously, the leading terms in the error can in general not be canceled exactly for the non-iterative case, as expressed by the tuning factor $\alpha$.
	The 
	\romannumeral 2.) The actual benchmark results discussed in Sec.~\ref{sec:co-simulation_benchmark_tests} show little to no dependence on this chance.
}, $m$ is the extrapolation order ($m=0$ for constant extrapolation), and $\alpha_\text{s} \in [0.8,0.9]$ is a safety factor.
The scalar error indicator can be defined as\cite{Sadjina2016}
\begin{equation}
\label{equ:error_indicator}
	\epsilon(t)
	\equiv
	\sqrt{
		\frac{1}{N}
		\sum_{k = 1}^N
		\bigg(
			\frac{
				\delta E_k(t)
			}{
				r_k
				\big(
				{E_0}_k
				+
				| E_k(t) |
				\big)
		}
		\bigg)^2
	}
	,
\end{equation}
using the residual energies $\delta E_k$ and energies $E_k(t_{i+1}) \approx P_{k_{12}}(t_{i+1}) \Delta t_i$ transmitted per time step for all $N$ power bonds.
Here, the typical energy scale ${E_0}_k$ and the relative tolerance $r_k$ are freely configurable parameters which determine the energy resolution for the power bond $k$.
The I-controller~\eqref{equ:step_control} aims to find and maintain a balance between accuracy and efficiency by choosing a step size for which $\epsilon \approx 1$:
Efficiency can be improved if $\epsilon < 1$ by increasing the step size, while accuracy needs to be increased by choosing smaller time steps of $\epsilon > 1$.
In order to avoid rapid oscillations in the step size on one hand, and inefficiently small step sizes on the other, the step size itself and its rate of change are restricted by the parameters ${\Delta t}_\text{min}$ and ${\Delta t}_\text{max}$, and $\Theta_\text{min}$ and $\Theta_\text{max}$, respectively.
Table~\ref{tab:controller_configuration} lists the full configuration used for the benchmark tests of Section~\ref{sec:co-simulation_benchmark_tests}.

\begin{table}[h!tb]
	\begin{tabular}{lS[table-format=3.1]s}
		\hline\noalign{\smallskip}
		 & {Value} & {Unit} \\
		\noalign{\smallskip}\hline\noalign{\smallskip}
		$\alpha_\text{s}$ & 0.8 & \\
		${\Delta t}_\text{min}$ & 10 & \micro\second \\
		${\Delta t}_\text{max}$ & 10 & \milli\second \\
		$\Theta_\text{min}$ & 0.2 & \\
		$\Theta_\text{max}$ & 1.5 & \\
		$E_0$ & 750 & \si{\joule} \\
		\noalign{\smallskip}\hline
	\end{tabular}
	\caption{%
		Configuration of the adaptive step size controller for the benchmark model in Sec.~\ref{sec:co-simulation_benchmark_tests}
	}
	\label{tab:controller_configuration}
\end{table}

%------------------------------------------------------------ 
%------------------------------------------------------------ 
\section{Co-Simulation Benchmark Tests}
\label{sec:co-simulation_benchmark_tests}

\begin{figure}[h!tb]
	\centering
	\def\svgwidth{\graphicswidth}
	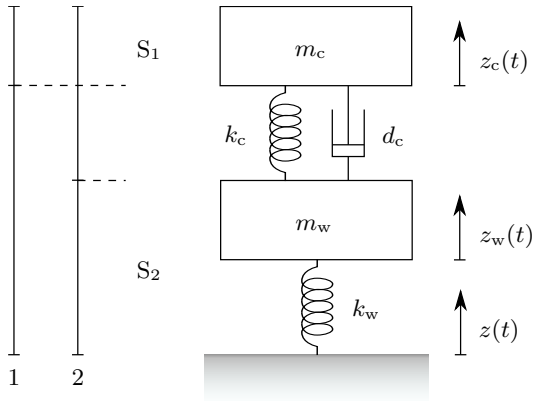
	\caption{%
		The quarter car benchmark model is split into the subsystems S$_1$ and S$_2$ for co-simulation using the two distinct reticulations 1 and 2
	}
	\label{fig:quartercar_model}
\end{figure}

In order to assess the performance of the methods discussed in sections~\ref{sec:energy_conserving_corrections} and~\ref{sec:adaptive_step_size}, we employ a quarter car model as described in Ref.~\onlinecite{Arnold2013} and split it into two subsystems connected via a power bond, see Fig.~\ref{fig:quartercar_model}.
This model can be considered two coupled Dahlquist test equations\cite{Dahlquist1956} and is thus well suited as a co-simulation benchmark test case.\cite{Clauss2012,Schierz2012,Arnold2013,Arnold2014,Sadjina2016}
We further study two different reticulations for the co-simulation and also investigate nonlinear damping characteristics.
The corresponding model and the underlying equations are adapted directly from Ref.~\onlinecite{Sadjina2016}, the parameters are summarized in Table~\ref{tab:quartercar_model} for the linear test case and in Table~\ref{tab:quartercar_nonlinear_model} for the nonlinear case.

\begin{table}[h!tb]
	\centering
	\begin{tabular}{lS[table-format=6.1]s}
		\hline\noalign{\smallskip}
		 & {Value} & Unit \\
		\noalign{\smallskip}\hline\noalign{\smallskip}
		$m_\text{c}$ & 400 & \kilogram \\
		$m_\text{w}$ & 40 & \kilogram \\
		$k_\text{c}$ & 15000 & \newton\per\meter \\
		$k_\text{w}$ & 150000 & \newton\per\meter \\
		$d_\text{c}$ & 1000 & \newton\second\per\meter \\
		$n_d$ & 0.5 & \\
		\noalign{\smallskip}\hline
	\end{tabular}
	\caption{%
		Parameters for the linear quarter car benchmark model according to Ref.~\onlinecite{Arnold2013}
	}
	\label{tab:quartercar_model}
\end{table}

\begin{table}[h!tb]
	\centering
	\begin{tabular}{lS[table-format=3.1]s}
		\hline\noalign{\smallskip}
		 & {Value} & {Unit} \\
		\noalign{\smallskip}\hline\noalign{\smallskip}
		$d_\text{c}$ & 900 & \newton\raiseto{1/2}{(\second\per\meter)} \\
		$n_d$ & 1.5 & \\
		\noalign{\smallskip}\hline
	\end{tabular}
	\caption{%
		Parameter changes to include nonlinear damping forces in the benchmark model according to Ref.~\onlinecite{Busshardt1992}
	}
	\label{tab:quartercar_nonlinear_model}
\end{table}

We generally carry out the time integrations in the subsimulators using micro step sizes of ${\Delta t}_{\text{S}_1} = {\Delta t}_{\text{S}_2} = {\Delta t}/256$ with the forward Euler method to focus on the co-simulation coupling errors.\footnote{%
	Even smaller micro step sizes affect the benchmark results only marginally.
}
As mentioned previously, we use energies and powers as error metrics to assess the quality of the co-simulation results:
On one hand, we consider the average error in the power~\eqref{equ:energytransfer} transmitted over the power bond from simulator S$_1$ to simulator S$_2$,
\begin{subequations}
\label{equ:error_measures}
\begin{equation}
\label{equ:error_measures_power}
	\Delta P(t_{i+1})
	\equiv
	\frac{1}{T}
	\sum_{j=0}^i
	|
		P_{12}(t_{j+1}) - P^0_{12}(t_{j+1})
	|
	\Delta t_j,
\end{equation}
where $P^0_{12}(t)$ is the exact solution and $T$ is the total (virtual) duration of the simulation run.
On the other hand, the total accumulated residual energy
\begin{equation}
\label{equ:error_measures_total_residual_energy}
	\Delta E(t_{i+1})
	\equiv
	\sum_{j=0}^i
	\delta P(t_{j+1})
	\Delta t_j
\end{equation}
\end{subequations}
gives the amount of energy wrongfully added to the full system during the entire simulation time interval $t \in [t_0, t_{i+1}]$ and is thus used as another indicator of co-simulation accuracy.

%------------------------------------------------------------ 
\subsection{NEPCE}
\label{subsec:co-simulation_benchmark_tests_NEPCE}

Let us first use the quarter car model to benchmark NEPCE's performance.
The tuning factor $\alpha$ is chosen such that the errors are minimized while avoiding the excitation of fast oscillations and risking instability.
The energy errors can be reduced throughout by \SI{49}{\percent} to \SI{86}{\percent} when using NEPCE compared to the uncorrected results.
The results are summarized in Tables~\ref{tab:quartercar_NEPCE} and~\ref{tab:quartercar_alt_NEPCE}, where the tuning factor, the total number of macro time steps, and the power transmitted over the power bond $P_{12}$ averaged over the entire simulation duration $T$ are shown.
Furthermore, the error in the power $\Delta P(T)$ and the total accumulated residual energy $\Delta E(T)$ are given according to Eqs.~\eqref{equ:error_measures} with respect to the simulation duration $T$.

\begin{table}[h!tb]
\centering
	\begin{tabular}{lS[table-format=1.2]S[table-format=4.0]S[table-format=1.2]S[table-format=1.2]S[table-format=1.2]}
		\hline\noalign{\smallskip}
  		\multicolumn{3}{c}{Algorithm} & \multicolumn{1}{ c }{Power} & \multicolumn{2}{ c }{Error} \\
 		\multicolumn{1}{ c }{type} & \multicolumn{1}{ c }{tuning} & \multicolumn{1}{ c }{steps} & \multicolumn{1}{ c }{$\frac{\overline{P_{12}}}{\si{\watt}}$} & \multicolumn{1}{ c }{$\frac{\Delta P}{\si{\watt}}$} & \multicolumn{1}{ c}{$\frac{\Delta E}{\si{\joule}}$} \\
		\noalign{\smallskip}\hline\noalign{\smallskip}
		constant & & 4000 & 0.4 & 1.0 & 6.3 \\
		\noalign{\smallskip}\hline\noalign{\smallskip}
		NEPCE & 0.95 & 4000 & 0.01 & 0.14 & 3.20 \\
		\noalign{\smallskip}\hline\noalign{\smallskip}
		NEPCE mod.\ & 0.95 & 4000 & 0.01 & 0.11 & 3.20 \\
		\noalign{\smallskip}\hline
	\end{tabular}
	\caption{%
		Linear quarter car benchmark results for reticulation 1 with NEPCE and with NEPCE with direct feed-through modification
	}
	\label{tab:quartercar_NEPCE}
\end{table}

\begin{table}[h!tb]
\centering
	\begin{tabular}{lS[table-format=1.2]S[table-format=4.0]S[table-format=-1.2]S[table-format=1.2]S[table-format=1.2]}
		\hline\noalign{\smallskip}
  		\multicolumn{3}{c}{Algorithm} & \multicolumn{1}{ c }{Power} & \multicolumn{2}{ c }{Error} \\
 		\multicolumn{1}{ c }{type} & \multicolumn{1}{ c }{tuning} & \multicolumn{1}{ c }{steps} & \multicolumn{1}{ c }{$\frac{\overline{P_{12}}}{\SI{d2}{\watt}}$} & \multicolumn{1}{ c }{$\frac{\Delta P}{\SI{d2}{\watt}}$} & \multicolumn{1}{ c}{$\frac{\Delta E}{\SI{d2}{\joule}}$} \\
		\noalign{\smallskip}\hline\noalign{\smallskip}
		constant & & 4000 & -1.89 & 0.10 & 0.22 \\
		\noalign{\smallskip}\hline\noalign{\smallskip}
		NEPCE & 0.85 & 4000 & -1.88 & 0.04 & 0.11 \\
		\noalign{\smallskip}\hline\noalign{\smallskip}
		NEPCE mod.\ & 0.85 & 4000 & -1.88 & 0.03 & 0.10 \\
		\noalign{\smallskip}\hline
	\end{tabular}
	\caption{%
		Linear quarter car benchmark results for reticulation 2 with NEPCE and with NEPCE with direct feed-through modification
	}
	\label{tab:quartercar_alt_NEPCE}
\end{table}

\begin{figure}[h!tb]
	\centering
	\includegraphics[width=\graphicswidth]{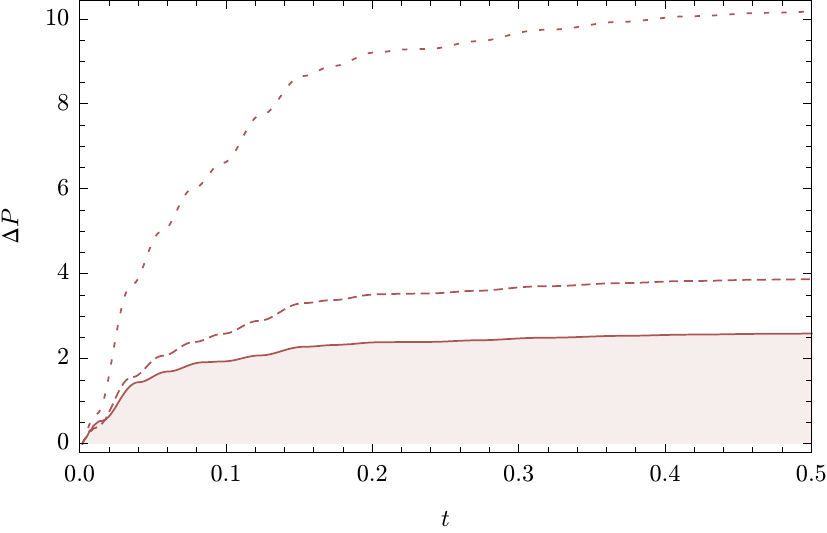}
	\caption{%
		Average error in the power for the linear quarter car benchmark with reticulation 2 and constant step size:
		NEPCE with direct feed-through modification (solid), NEPCE alone (dashed), and the uncorrected result (dotted)
	}
	\label{fig:quartercar_comparisonNEPCE}
\end{figure}

The quarter car benchmark model does exhibit direct feed-through (in simulator S$_2$ for system reticulation 1 and in S$_1$ in system reticulation 2).
We thus expect improved performance when including the modification to NEPCE discussed in Section~\ref{subsec:corrections_feedthrough}.
Indeed, the average error in the power $\Delta P(T)$ is reduced by about another \SI{17}{\percent} to \SI{33}{\percent} with the modification.
Fig.~\ref{fig:quartercar_comparisonNEPCE} exemplifies this enhancement by showing the average error in the transmitted power for system reticulation 2.
Note, however, that the direct feed-through modification to NEPCE does not significantly influence the overall accumulated residual energy $\Delta E(T)$.

%------------------------------------------------------------ 
\subsection{NEPCE with ECCO}
\label{subsec:co-simulation_benchmark_tests_ECCO}

Let us now demonstrate how the corrections to the inputs are made more efficient by combining them with ECCO's energy-conservation-based adaptive step size control, as proposed in Sec.~\ref{sec:adaptive_step_size}.
For this purpose, the I controller~\eqref{equ:step_control} and the scalar error indicator~\eqref{equ:error_indicator} are configured according to the parameters listed in Table~\ref{tab:controller_configuration}, and the starting step size is set to ${\Delta t}_0 = {\Delta t}_\text{min}$.
The quarter car system is initially excited with an energy of \SI{750}{\joule} which thus determines the characteristic energy scale for the system, $E_0 = \SI{750}{\joule}$.
The tolerance $r$ is set such that the total number of macro time steps remains around a constant \num{4000} steps in order to keep the computational cost at the same level.

\begin{figure}[h!tb]
	\centering
	\includegraphics[width=\graphicswidth]{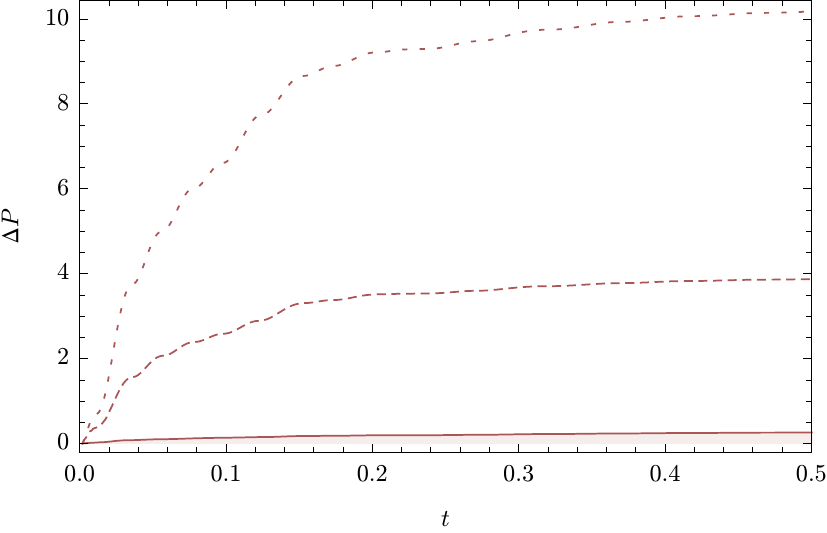}
	\caption{%
		Average error in the power for the linear quarter car benchmark with reticulation 2:
		ECCO with NEPCE (solid) against the constant step size results with NEPCE (dashed) and without any corrections (dotted)
	}
	\label{fig:quartercar_comparisonECCO}
\end{figure}

Substantial improvements are observed when using NEPCE with ECCO's adaptive step size control:
The energy errors in the benchmarks are reduced by \SI{87}{\percent} to \SI{92}{\percent} for system reticulation 1, see Table~\ref{tab:quartercar_ECCO}, and by \SI{97}{\percent} to \SI{98}{\percent} for system reticulation 2, see Table~\ref{tab:quartercar_alt_ECCO}.
This considerable enhancement of the quality of the co-simulation results is also exemplified in Fig.~\ref{fig:quartercar_comparisonECCO}.

\begin{table}[h!tb]
\centering
	\begin{tabular}{lS[table-format=1.2]S[table-format=1.1d-1]S[table-format=4.0]S[table-format=-1.2]S[table-format=1.2]S[table-format=1.2]}
		\hline\noalign{\smallskip}
  		\multicolumn{4}{c}{Algorithm} & \multicolumn{1}{ c }{Power} & \multicolumn{2}{ c }{Error} \\
 		\multicolumn{1}{ c }{type} & \multicolumn{1}{ c }{tuning} & \multicolumn{1}{ c }{tolerance} & \multicolumn{1}{ c }{steps} & \multicolumn{1}{ c }{$\frac{\overline{P_{12}}}{\si{\watt}}$} & \multicolumn{1}{ c }{$\frac{\Delta P}{\si{\watt}}$} & \multicolumn{1}{ c}{$\frac{\Delta E}{\si{\joule}}$} \\
		\noalign{\smallskip}\hline\noalign{\smallskip}
		constant & & & 4000 & 0.4 & 1.0 & 6.3 \\
		\noalign{\smallskip}\hline\noalign{\smallskip}
		NEPCE & 0.95 & 1.6d-6 & 3930 & -0.05 & 0.08 & 0.83 \\
		\noalign{\smallskip}\hline\noalign{\smallskip}
		NEPCE mod.\ & 0.95 & 1.6d-6 & 4002 & -0.04 & 0.06 & 0.81 \\
		\noalign{\smallskip}\hline
	\end{tabular}
	\caption{%
		Linear quarter car benchmark results for reticulation 1 using residual-energy-based adaptive step size control with NEPCE and with NEPCE with direct feed-through modification
	}
	\label{tab:quartercar_ECCO}
\end{table}

\begin{table}[h!tb]
\centering
	\begin{tabular}{lS[table-format=1.2]S[table-format=1.1d-1]S[table-format=4.0]S[table-format=-1.3]S[table-format=1.3]S[table-format=1.3]}
		\hline\noalign{\smallskip}
  		\multicolumn{4}{c}{Algorithm} & \multicolumn{1}{ c }{Power} & \multicolumn{2}{ c }{Error} \\
 		\multicolumn{1}{ c }{type} & \multicolumn{1}{ c }{tuning} & \multicolumn{1}{ c }{tolerance} & \multicolumn{1}{ c }{steps} & \multicolumn{1}{ c }{$\frac{\overline{P_{12}}}{\SI{d2}{\watt}}$} & \multicolumn{1}{ c }{$\frac{\Delta P}{\SI{d2}{\watt}}$} & \multicolumn{1}{ c}{$\frac{\Delta E}{\SI{d2}{\joule}}$} \\
		\noalign{\smallskip}\hline\noalign{\smallskip}
		constant & & & 4000 & -1.89 & 0.10 & 0.22 \\
		\noalign{\smallskip}\hline\noalign{\smallskip}
		NEPCE & 0.85 & 1.4d-6 & 3921 & -1.872 & 0.003 & 0.004 \\
		\noalign{\smallskip}\hline\noalign{\smallskip}
		NEPCE mod.\ & 0.85 & 1.4d-6 & 3958 & -1.871 & 0.002 & 0.004 \\
		\noalign{\smallskip}\hline
	\end{tabular}
	\caption{%
		Linear quarter car benchmark results for reticulation 2 using residual-energy-based adaptive step size control with NEPCE and with NEPCE with direct feed-through modification
	}
	\label{tab:quartercar_alt_ECCO}
\end{table}

\begin{figure}[h!tb]
	\centering
	\includegraphics[width=\graphicswidth]{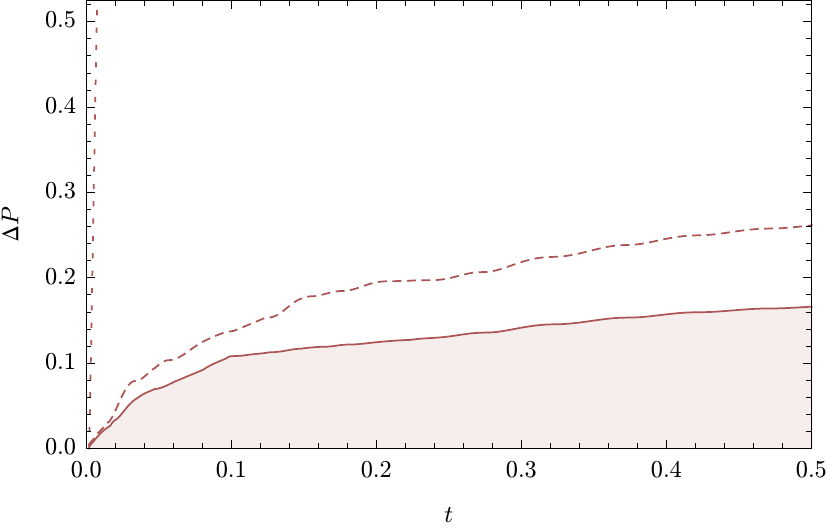}
	\caption{%
		Average error in the power for the linear quarter car benchmark with reticulation 2:
		ECCO with NEPCE with direct feed-through modification (solid), ECCO with NEPCE alone (dashed), and uncorrected result with constant step sizes (dotted)
	}
	\label{fig:quartercar_comparisonECCO_mod_NEPCE}
\end{figure}

The situation is further improved by also including the direct feed-through modifications for NEPCE, as shown in Fig.~\ref{fig:quartercar_comparisonECCO_mod_NEPCE}.
Then, an additional reduction of the average error in the power of \SI{26}{\percent} to \SI{36}{\percent} is achieved compared to the results without the modification.
Again, however, the accumulated residual energy is almost unaffected.

In conclusion, the non-iterative energy-conservation-based co-simulation framework presented here (NEPCE with direct feed-through modification and ECCO) manages to reduce the energy errors by between \SI{87}{\percent} and \SI{98}{\percent} in the linear quarter car benchmark at no additional computational cost.

%------------------------------------------------------------ 
\subsection{Nonlinear Damping}
\label{subsec:nonlinear_spring}

Finally, let us study the effects of nonlinear damping as given in Table~\ref{tab:quartercar_nonlinear_model}.
Note that the total simulation duration is now set to $T = \SI{2}{\second}$ (\num{2000} macro time steps in total) because the excitations in the system are subdued faster with the more efficient nonlinear damper.
In addition, system reticulation 2 is relatively unstable for nonlinear damping, and the macro step size is thus restricted to $t_\text{max} = \SI{2.5}{\milli\second}$ for this setup.

The energy-conservation-based corrections to the inputs (as expressed by the tuning factor $\alpha$) have to be applied less aggressively to avoid rapid oscillations.
Yet, using NEPCE alone without modifications yields a reduction in the energy errors of \SI{32}{\percent} to \SI{60}{\percent} when  compared to the uncorrected results, as shown in Tables~\ref{tab:quartercar_nonlinear} and~\ref{tab:quartercar_alt_nonlinear}.
As was the case for the linear benchmark, significant improvements are obtained by combining NEPCE with ECCO:
The energy errors are reduced by \SI{79}{\percent} to \SI{91}{\percent} compared to uncorrected results obtained with a constant step size.
Also including the direct feed-through modifications with NEPCE leads to small additional reductions of \SI{0}{\percent} to \SI{19}{\percent}.

\begin{table}[h!tb]
\centering
	\begin{tabular}{lS[table-format=1.1]S[table-format=1.1d-1]S[table-format=4.0]S[table-format=-1.1]S[table-format=1.1]S[table-format=1.1]}
		\hline\noalign{\smallskip}
  		\multicolumn{4}{c}{Algorithm} & \multicolumn{1}{ c }{Power} & \multicolumn{2}{ c }{Error} \\
 		\multicolumn{1}{ c }{type} & \multicolumn{1}{ c }{tuning} & \multicolumn{1}{ c }{tolerance} & \multicolumn{1}{ c }{steps} & \multicolumn{1}{ c }{$\frac{\overline{P_{12}}}{\si{\watt}}$} & \multicolumn{1}{ c }{$\frac{\Delta P}{\si{\watt}}$} & \multicolumn{1}{ c}{$\frac{\Delta E}{\si{\joule}}$} \\
		\noalign{\smallskip}\hline\noalign{\smallskip}
		constant & & & 2000 & 0.6 & 1.3 & 4.7 \\
		\noalign{\smallskip}\hline\noalign{\smallskip}
		NEPCE & 0.6 &  & 2000 & 0.1 & 0.5 & 2.9 \\
		\noalign{\smallskip}\hline\noalign{\smallskip}
		NEPCE mod.\ & 0.6 &  & 2000 & 0.1 & 0.5 & 2.9 \\
		\noalign{\smallskip}\hline\noalign{\smallskip}
		NEPCE & 0.6 & 4.7d-6 & 1991 & -0.1 & 0.2 & 1.0 \\
		\noalign{\smallskip}\hline\noalign{\smallskip}
		NEPCE mod.\ & 0.6 & 4.6d-6 & 2010 & -0.1 & 0.2 & 1.0 \\
		\noalign{\smallskip}\hline
	\end{tabular}
	\caption{%
		Nonlinear quarter car benchmark results for reticulation 1
	}
	\label{tab:quartercar_nonlinear}
\end{table}

\begin{table}[h!tb]
\centering
	\begin{tabular}{lS[table-format=1.1]S[table-format=1.1d-1]S[table-format=4.0]S[table-format=-1.2]S[table-format=1.2]S[table-format=1.2]}
		\hline\noalign{\smallskip}
  		\multicolumn{4}{c}{Algorithm} & \multicolumn{1}{ c }{Power} & \multicolumn{2}{ c }{Error} \\
 		\multicolumn{1}{ c }{type} & \multicolumn{1}{ c }{tuning} & \multicolumn{1}{ c }{tolerance} & \multicolumn{1}{ c }{steps} & \multicolumn{1}{ c }{$\frac{\overline{P_{12}}}{\SI{d2}{\watt}}$} & \multicolumn{1}{ c }{$\frac{\Delta P}{\SI{d2}{\watt}}$} & \multicolumn{1}{ c}{$\frac{\Delta E}{\SI{d2}{\joule}}$} \\
		\noalign{\smallskip}\hline\noalign{\smallskip}
		constant & & & 2000 & -3.8 & 0.2 & 0.4 \\
		\noalign{\smallskip}\hline\noalign{\smallskip}
		NEPCE & 0.4 &  & 2000 & -3.78 & 0.14 & 0.30 \\
		\noalign{\smallskip}\hline\noalign{\smallskip}
		NEPCE mod.\ & 0.4 &  & 2000 & -3.78 & 0.12 & 0.30 \\
		\noalign{\smallskip}\hline\noalign{\smallskip}
		NEPCE & 0.4 & 2.6d-5 & 1986 & -3.77 & 0.04 & 0.04 \\
		\noalign{\smallskip}\hline\noalign{\smallskip}
		NEPCE mod.\ & 0.4 & 2.7d-5 & 1989 & -3.77 & 0.03 & 0.04 \\
		\noalign{\smallskip}\hline
	\end{tabular}
	\caption{%
		Nonlinear quarter car benchmark results for reticulation 2
	}
	\label{tab:quartercar_alt_nonlinear}
\end{table}

%------------------------------------------------------------ 
%------------------------------------------------------------ 
\section{Conclusion}
\label{sec:conclusion}

The \emph{Energy-Conservation-based Co-Simulation} method\cite{Sadjina2016} (ECCO) provides a generic framework for error estimation and adaptive step size control in non-iterative co-simulations.
Using power bonds to realize the simulator coupling, it directly monitors power flows between the subsystems and gives the exact amount of energy wrongfully added to the total energy of the full coupled system during co-simulation (macro) time steps.
The resulting so-called \emph{residual energies} are obtain from the coupling variable values alone, and ECCO uses them to propose an optimal macro time step to minimize energy errors throughout the co-simulation.
The \emph{Nearly Energy Preserving Coupling Element}\cite{Benedikt2013} (NEPCE), on the other hand, corrects for coupling errors in non-iterative co-simulations directly to make the flow of (generalized) energy between subsimulators more accurate.

In the present paper, we combine both methods to optimize the efficiency and accuracy of non-iterative co-simulations.
NEPCE is based on the assumption that the coupling variables are slowly varying on the scale of the macro time step.
ECCO, on the other hand, provides a systematic approach to fulfill this requirement by adaptively controlling the macro step size in order to minimize the violation of energy conservation.
We also extend NEPCE to the case where direct feed-through is present.
Then, the output errors give contributions to the residual energy which are of the same order as the ones stemming from the input errors.
Put differently, additional contributions to the violation of energy conservation should be taken into account when constructing energy-conserving corrections to the coupling variables.
This is, however, only possible if the interface Jacobian is known.

The performance of the concepts discussed here is demonstrated by use of a quarter car co-simulation benchmark model.
We study two distinct system reticulations, as well as the effects of including nonlinear damping characteristics.
In these benchmarks, NEPCE alone generally yields a reduction in the energy errors of \SI{32}{\percent} to \SI{86}{\percent}, depending on how aggressively it can be used before unwanted oscillations are induced.
The proposed direct feed-through modification to NEPCE reduces the energy errors by another \SI{0}{\percent} to \SI{36}{\percent}.
Also employing ECCO's adaptive step size control leads to substantially higher accuracies in the co-simulation results:
Energy errors are then reduced by up to \SI{98}{\percent} when compared to the uncorrected results with constant macro step sizes.

%------------------------------------------------------------ 
%------------------------------------------------------------ 
\begin{acknowledgments}

This work was funded by the Research Council of Norway (project no. $225322$ MAROFF) and the industrial partners in the ViProMa project consortium (VARD, Rolls-Royce Marine and DNV GL).
We are grateful for their financial support.
The authors would further like to thank Stian Skjong for fruitful discussions.

\end{acknowledgments}

%------------------------------------------------------------ 
%------------------------------------------------------------ 
\nocite{*}    
\bibliographystyle{ieeetr}
\bibliography{energycorrections}

\end{document}